\documentclass[12pt]{article}

\usepackage{epsf}
\usepackage[dvips]{graphicx}

\begin{document}
\begin{center}

{\bfseries NUCLEAR SPIN IN DIRECT DARK MATTER SEARCH}\vskip 5mm

V.A. Bednyakov$^{1 \dag}$, F. \v Simkovic$^{2}$ 
and I.V. Titkova$^{1}$ \vskip 5mm

{\small (1) 
{\it  Joint Institute for Nuclear Research, 
      Dzhelepov Laboratory of Nuclear Problems,
      141980 Dubna, Moscow Region, Russia}, \
       $\dag$ {\it E-mail: bedny@jinr.ru}  \\
(2) {\it Department of Nuclear Physics, Comenius University,
     Mlynsk\'a dolina F1, SK--842 15 Bratislava, Slovakia}}
\end{center}\vskip 5mm

\begin{center}
\begin{minipage}{150mm}
\centerline{\bf Abstract}
        The Weakly Interacting Massive Particles (WIMPs)
        are among the main candidates for the relic dark matter (DM).
	The idea of the direct DM detection relies on elastic 
	spin-dependent (SD) and spin-independent (SI) 
	interaction of WIMPs with target nuclei. 
	The importance of the SD WIMP-nucleus 
	interaction for reliable DM detection is argued. 
 	The absolute lower bound for the detection rate 
	can naturally be due to SD interaction.
	An experiment aimed at {\em detecting}\ DM 
	with sensitivity higher than $10^{-5}\,$event$/$day$/$kg
	should have a non-zero-spin target.
\end{minipage}
\end{center}\vskip 10mm

\section{Introduction}
        The lightest supersymmetric (SUSY) particle (LSP) neutralino is 
        assumed to be the Weakly Interacting Massive Particle (WIMP) 
	and the best dark matter (DM) candidate.  
        It is believed that for heavy enough nuclei
	this spin-independent (SI) interaction of 
        DM particles with nuclei usually makes the 
	dominant contribution to the expected event rate of its detection.
        The reason is the strong (proportional to the squared mass 
        of the target nucleus) 
        enhancement of SI WIMP-nucleus interaction.
	Nevertheless 
        there are at least three reasons to think that 
	SD (or axial-vector) interaction of the DM WIMPs with nuclei
        could be very important. 
        First, contrary to the only one constraint for SUSY models available 
        from the scalar WIMP-nucleus interaction, the spin WIMP-nucleus 
        interaction supplies us with two such constraints (see for example 
\cite{Bednyakov:1994te} and formulas below).
        Second, one can notice 
\cite{Bednyakov:2000he,Bednyakov:2002mb}
        that even with a very sensitive DM detector
        (say, with a sensitivity of $10^{-5}\,$events$/$day$/$kg)
        which is sensitive only to the WIMP-nucleus 
        scalar interaction (with spin-less target nuclei) 
        one can, in principle, miss a DM signal. 
        To safely avoid such a situation one should
        have a spin-sensitive DM detector, i.e. a detector 
        with non-zero-spin target nuclei.
        Finally, there is a complicated 
        nuclear spin structure, which, for example, 
	characterized by 
        the so-called long $q$-tail form-factor behavior. 
        Therefore for heavy target nuclei and heavy WIMP
	the SD efficiency to detect a DM signal 
        is much higher than the SI efficiency
\cite{Engel:1991wq}.  
	However, simultaneous study of both 
        spin-dependent and spin-independent interactions of the
        DM particles with nuclei significantly increases 
	the chance to observe the DM signal. 

\section{Two constrainsts for SUSY due to the spin}
\begin{figure}[t]
\begin{picture}(100,100) 
\put(22,-15){\includegraphics{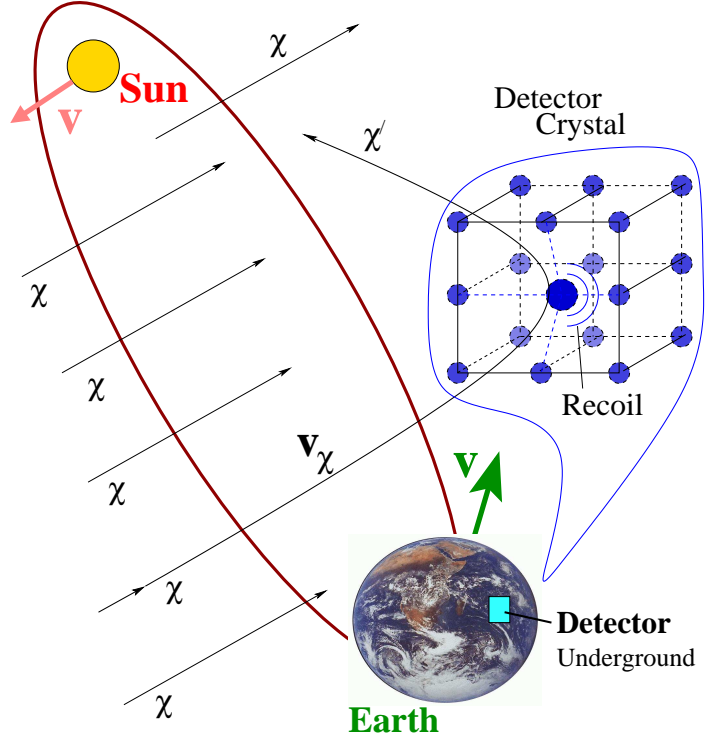}}
\end{picture} 
\caption{Due to the expected annual modulation signature of the event rate 
  (\ref{Definitions.diff.rate}) only the Sun-Earth system 
  is a proper setup for the successful direct DM detection.
\label{DDDDD}}
\end{figure}
        One believes to detect directly a DM particle $\chi$ 
	via its elastic scattering on a target nucleus $(A,Z)$. 
	The nuclear recoil energy 
	$E_{\rm R}$ 
	($E_{\rm R}\sim 10^{-6} m_{\chi} \approx$ few keV) 
	is measured by a proper detector 
(Fig.~\ref{DDDDD}).
	The differential event rate 
	depends on the distribution of
        the relic DM particles in the solar vicinity $f(v)$ and
        the cross section of LSP-nucleus elastic scattering
\cite{Jungman:1996df}--\cite{Bednyakov:1994qa}:
\begin{equation}
\label{Definitions.diff.rate}
	\frac{dR}{dE_{\rm R}} = N \frac{\rho_\chi}{m_\chi}
	\displaystyle
	\int^{v_{\max}}_{v_{\min}} dv f(v) v
	{\frac{d\sigma}{dq^2}} (v, q^2),
	\qquad
	E_{\rm R} = q^2 /(2 M_A ).  
\label{eq:1}
\end{equation}
	Here, $N={\cal N}/A$ is the number density 
	of target nuclei. ${\cal N}$ and  $A$ stand for 
	the Avogadro number and the atomic mass in AMU, respectively.   
        $M_A$ denotes the nuclear mass.
	$v_{\max} = v_{\rm esc} \approx 600$~km/s, \ 
	$v_{\min}=\left(M_A E_{\rm R}/2 \mu_{A}^2\right)^{1/2}$,
        the DM density
        $\rho_{\chi}$ = 0.3 GeV$\cdot$cm$^{-3}$. 
	The neutralino-nucleus elastic scattering cross section 
	for spin-non-zero ($J\neq 0$) nuclei is a sum of
        the coherent (spin-independent) and axial 
	(spin-dependent) terms
\cite{Engel:1991wq,Engel:1992bf,Ressell:1993qm,Ressell:1997kx}:
\begin{eqnarray}
\nonumber
{\frac{d\sigma^{A}}{dq^2}}(v,q^2) 
	&=& \frac{\sum{|{\cal M}|^2}}{\pi\, v^2 (2J+1)} 
         =  \frac{S^A_{\rm SD} (q^2)}{v^2 (2J+1)} 
           +\frac{S^A_{\rm SI} (q^2)}{v^2 (2J+1)} \\
\label{Definitions.cross.section}
        &=& \frac{\sigma^A_{\rm SD}(0)}{4\mu_A^2 v^2}F^2_{\rm SD}(q^2)
           +\frac{\sigma^A_{\rm SI}(0)}{4\mu_A^2 v^2}F^2_{\rm SI}(q^2).
\label{eq:2}
\end{eqnarray}
        It is useful to separate the zero-momentum transfer cross sections
	and introduce the normalized-to-unity ($F^2_{\rm SD,SI}(0) = 1$)
	nonzero-momentum-transfer nuclear form-factors:
\begin{equation}
\label{Definitions.form.factors}
F^2_{\rm SD,SI}(q^2) = \frac{S^{A}_{\rm SD,SI}(q^2)}{S^{A}_{\rm SD,SI}(0)} 
\label{eq:3}.
\end{equation}
	The SD structure function 
	$S^A_{\rm SD}(q)$ contains 
	the isoscalar $S_{00}$, isovector $S_{11}$ 
	and the interference $S_{01}$ terms: 
\begin{equation}
S^A_{\rm SD}(q) = a_0^2 S_{00}(q) + a_1^2 S_{11}(q) + a_0 a_1 S_{01}(q).
\label{Definitions.spin.decomposition} 
\end{equation}
	Here the isoscalar $a_0 = a_n + a_p$ and isovector 
	$a_1 = a_p - a_n$ effective coupling constants are used (see 
	(\ref{a_pn})). 
	For $q=0$  the nuclear SD and SI cross sections  
	take the forms  
\begin{eqnarray}
\label{NuclCS0}
\sigma^A_{\rm SI}(0) 
	&=& \frac{4\mu_A^2 \ S^{}_{\rm SI}(0)}{(2J+1)}\! =\!
	     \frac{\mu_A^2}{\mu^2_p}A^2 \sigma^{p}_{{\rm SI}}(0), \\ 
\sigma^A_{\rm SD}(0)
	&=&  \frac{4\mu_A^2 S^{}_{\rm SD}(0)}{(2J+1)}\! =\!
	     \frac{4\mu_A^2}{\pi}\frac{(J+1)}{J}
             \left\{a_p\langle {\bf S}^A_p\rangle 
                  + a_n\langle {\bf S}^A_n\rangle\right\}^2 . 
\label{NuclCS1}
\end{eqnarray}
	Here, $\displaystyle \mu_A = \frac{m_\chi M_A}{m_\chi+ M_A}$
	is the reduced $\chi$-nucleus mass
	and $\mu^2_{n}=\mu^2_{p}$ is assumed.
	The dependence on effective neutralino-quark couplings 
	${\cal C}_{q}$ and ${\cal A}_{q}$ in the underlying (SUSY) theory
\begin{equation}
{\cal  L}_{eff} = \sum_{q}^{}\left( 
	{\cal A}_{q}\cdot
      \bar\chi\gamma_\mu\gamma_5\chi\cdot
                \bar q\gamma^\mu\gamma_5 q + 
	{\cal C}_{q}\cdot\bar\chi\chi\cdot\bar q q
	\right)
      \ + ... 
\label{eq:6}
\end{equation}
	and on the spin ($\Delta^{(p,n)}_q$)
	and the mass ($f^{(p,n)}_q$) structure of {\em nucleons}\ 
	enter into these formulas via the zero-momentum-transfer 
	proton and neutron SI and SD cross sections: 
\begin{eqnarray}
\sigma^{p}_{{\rm SI}}(0) 
	= 4 \frac{\mu_p^2}{\pi}c_{0}^2,
&\qquad&
\sigma^{p,n}_{{\rm SD}}(0)  
 	=  12 \frac{\mu_{p,n}^2}{\pi}{a}^2_{p,n}; \\
	c^{p,n}_0 = \sum_q {\cal C}_{q} f^{(p,n)}_q,
&\quad&
	a_p =\sum_q {\cal A}_{q} \Delta^{(p)}_q, \quad 
	a_n =\sum_q {\cal A}_{q} \Delta^{(n)}_q.
\label{a_pn}
\end{eqnarray}
	The factors $\Delta_{q}^{(p,n)}$, which parametrize the quark 
	spin content of the nucleon, are defined as
	$ \displaystyle 2 \Delta_q^{(n,p)} s^\mu 
	  \equiv 
          \langle p,s| \bar{\psi}_q\gamma^\mu \gamma_5 \psi_q    
          |p,s \rangle_{(p,n)}$.
	The $\langle {\bf S}^A_{p(n)} \rangle $ is the total 
	spin of protons 
	(neutrons) averaged over all $A$ nucleons of the nucleus $(A,Z)$:
\begin{equation}
\langle {\bf S}^A_{p(n)} \rangle 
     \equiv \langle A \vert  {\bf S}^A_{p(n)} \vert A \rangle 
     = \langle A \vert  \sum_i^A {\bf s}^i_{p(n)} \vert A \rangle 
\label{eq:9}
\end{equation}

        The mean velocity $\langle v \rangle$ of 
	the relic DM particles of our Galaxy is about  
	$300~{\rm km/s} = 10^{-3} c$.
	For not very heavy $m_{\chi}$ and $M_A$
	one can 
	use the SD matrix element in
	{\em zero momentum transfer limit}\
\cite{Ressell:1997kx,Engel:1995gw}
\begin{equation}
\label{Definitions.matrix.element}
 {\cal M} 
\propto \langle A\vert a_p {\bf S}_p + a_n {\bf S}_n
 	\vert A \rangle \cdot {\bf s}_{\chi}.
\label{eq:10}
\end{equation}
\begin{table}[p] 
\caption{Zero momentum spin structure of nuclei in different models. 
	The measured magnetic moments used as input 
	are enclosed in parentheses. 
From 
\cite{Bednyakov:2004xq}.
\label{Nuclear.spin.main.table}}
\begin{center}
\vspace*{-7pt}
\small
\begin{tabular}{lrrr}
\hline
\hline 
$^{19}$F~($L_J=S_{1/2}$) & 
~~~~~~~~$\langle {\bf S}_p \rangle$ & 
~~~~~~~~$\langle {\bf S}_n \rangle$ & ~~~~~~~~$\mu$ (in $\mu_N$) \\ \hline
ISPSM, Ellis--Flores~\cite{Ellis:1988sh,Ellis:1991ef}
	& $1/2$ 	& $0$ 	& $2.793$  \\ 
OGM, Engel--Vogel~\cite{Engel:1989ix} 
	& $0.46$ 	& $0$ 	& $(2.629)_{\rm exp}$ \\ 
EOGM ($g_A/g_V=1$), Engel--Vogel~\cite{Engel:1989ix} 
	& $0.415$ 	& $-0.047$ & $(2.629)_{\rm exp}$ \\  
EOGM ($g_A/g_V=1.25$), Engel--Vogel~\cite{Engel:1989ix} 
	& $0.368$ 	& $-0.001$ & $(2.629)_{\rm exp}$ \\ 
SM, Pacheco-Strottman~\cite{Pacheco:1989jz}
	& $0.441$   	&$-0.109$  \\
SM, Divari et al.~\cite{Divari:2000dc} 
	& $0.4751$ 	& $-0.0087$ 
	& $2.91$  \\  
\hline
\hline 
$^{23}$Na~($L_J=P_{3/2}$) & 
~~~~~~~~$\langle {\bf S}_p \rangle$ & 
~~~~~~~~$\langle {\bf S}_n \rangle$ & ~~~~~~~~$\mu$ (in $\mu_N$) \\ \hline
ISPSM 
	& $1/2$ 	& 0		& $3.793$  \\ 
SM, Ressell-Dean~\cite{Ressell:1997kx}
	& 0.2477 	& 0.0198 	& 2.2196 \\ 
OGM, Ressell-Dean~\cite{Ressell:1997kx}      
        & 0.1566 	& 0.0    & (2.218)$_{\rm exp}$ \\ 
SM, Divari ar al.~\cite{Divari:2000dc} 
	& 0.2477 & 0.0199 & 2.22   \\ 
\hline
\hline 
$^{27}$Al~($L_J=D_{5/2}$) & 
~~~~~~~~$\langle {\bf S}_p \rangle$ & 
~~~~~~~~$\langle {\bf S}_n \rangle$ & 
~~~~~~~~$\mu$ (in $\mu_N$) \\ \hline
ISPSM, Ellis--Flores~\cite{Ellis:1988sh,Ellis:1991ef}
	& $1/2$ 	& 0 	  & $4.793$  \\ 
OGM, Engel--Vogel~\cite{Engel:1989ix} 
	& $0.25$ 	& 0 	  & $(3.642)_{\rm exp}$ \\ 
EOGM ($g_A/g_V=1$), Engel--Vogel~\cite{Engel:1989ix} 
	& $0.333$ 	& $0.043$ & $(3.642)_{\rm exp}$ \\ 
EOGM ($g_A/g_V=1.25$), Engel--Vogel~\cite{Engel:1989ix} 
	&  $0.304$ 	& $0.072$ & $(3.642)_{\rm exp}$ \\ 
SM, Engel et al.~\cite{Engel:1995gw}
	& $0.3430$ & $0.0296$ 
	& $3.584$  \\
\hline\hline
$^{73}$Ge~($L_J=G_{9/2}$) & ~~~~~~~~$\langle {\bf S}_p \rangle$ & 
~~~~~~~~$\langle {\bf S}_n \rangle$ & ~~~~~~~~$\mu$ (in $\mu_N$) \\ \hline
ISPSM, Ellis--Flores~\cite{Ellis:1988sh,Ellis:1991ef}
	&    0	  & $0.5$		& $-1.913$ \\ 
OGM, Engel--Vogel~\cite{Engel:1989ix} 	
	&    0	  & $0.23$ 	&$(-0.879)_{\rm exp}$ \\ 
IBFM, Iachello et al.~\cite{Iachello:1991ut} and \cite{Ressell:1993qm}
	&$-0.009$ & $0.469$ &$-1.785$\\ 
IBFM (quenched), 
	Iachello et al.~\cite{Iachello:1991ut} and \cite{Ressell:1993qm}
	&$-0.005$  & $0.245$ &$(-0.879)_{\rm exp}$ \\
TFFS, Nikolaev--Klapdor-Kleingrothaus, \cite{Nikolaev:1993dd} 
	&$0$   & $0.34$ & --- \\ 
SM (small), Ressell et al.~\cite{Ressell:1993qm} 
	&$0.005$   & $0.496$ &$-1.468$ \\ 
SM (large), Ressell et al.~\cite{Ressell:1993qm} 
	&$0.011$   & $0.468$ &$-1.239$ \\ 
SM (large, quenched), Ressell et al.~\cite{Ressell:1993qm} 
	&$0.009$   & $0.372$ &$(-0.879)_{\rm exp}$ \\ 
``Hybrid'' SM, Dimitrov et al.~\cite{Dimitrov:1995gc}           
	& $0.030$ & $0.378$ & $-0.920$ \\ 
\hline
\hline
$^{127}$I~($L_J=D_{5/2}$) & 
~~~~~~~~$\langle {\bf S}_p \rangle$ & ~~~~~~~~$\langle {\bf S}_n \rangle$ & 
~~~~~~~~$\mu$ (in $\mu_N$) \\ \hline
ISPSM, Ellis--Flores~\cite{Ellis:1991ef,Ellis:1993vh}
	&$1/2$ & 0 &  $4.793$ \\ 
OGM, Engel--Vogel~\cite{Engel:1989ix} 
	&$0.07$ & 0 & $(2.813)_{\rm exp}$ \\ 
IBFM, Iachello et al.~\cite{Iachello:1991ut}
	&$0.464$ & $0.010$ & $(2.813)_{\rm exp}$ \\ 
IBFM (quenched), 
	Iachello et al.~\cite{Iachello:1991ut}
	&$0.154$  & $0.003$ & $(2.813)_{\rm exp}$ \\ 
TFFS, Nikolaev--Klapdor-Kleingrothaus, \cite{Nikolaev:1993dd} 
	& $0.15$ & 0 & --- \\ 
SM (Bonn A), Ressell--Dean~\cite{Ressell:1997kx} 
	& $0.309$ & $0.075$ & $2.775~\{2.470\}_{\rm eff}$  \\ 
SM (Nijmegen II), Ressell--Dean~\cite{Ressell:1997kx} 
	& $0.354$ & $0.064$ & $3.150~\{2.7930\}_{\rm eff}$ \\ 
\hline\hline
$^{131}$Xe~($L_J=D_{3/2}$) & 
~~~~~~~~$\langle {\bf S}_p \rangle$ & 
~~~~~~~~$\langle {\bf S}_n \rangle$ & 
~~~~~~~~$\mu$ (in $\mu_N$) \\ \hline 
ISPSM, Ellis--Flores~\cite{Ellis:1988sh,Ellis:1991ef}
	& 0 & $-0.3$ & $1.148$ \\
OGM, Engel--Vogel~\cite{Engel:1989ix} 
	& 0.0  & $-0.18$ 	& $(0.692)_{\rm exp}$ \\ 
IBFM, Iachello et al.~\cite{Iachello:1991ut}
	&$0.000$ & $-0.280$ 	& $(0.692)_{\rm exp}$ \\
IBFM (quenched), Iachello et al.~\cite{Iachello:1991ut}
	&$0.000$  & $-0.168$ 	& $(0.692)_{\rm exp}$ \\
TFFS, Nikolaev--Klapdor-Kleingrothaus, \cite{Nikolaev:1993dd} 
	&   & $-0.186$ & --- \\ 
SM (Bonn A), Ressell--Dean~\cite{Ressell:1997kx} 
	& $-0.009$ & $-0.227$ & $0.980~\{0.637\}_{\rm eff}$  \\ 
SM (Nijmegen II), Ressell--Dean~\cite{Ressell:1997kx} 
	& $-0.012$ & $-0.217$ & $0.979~\{0.347\}_{\rm eff}$ \\  
QTDA, Engel~\cite{Engel:1991wq} 
	& $-0.041$ & $-0.236$ & 0.70 \\  
\hline\hline
\end{tabular} \end{center}
\end{table} 

        Note a coupling  of the spin of $\chi$, 
	${\bf s}_{\chi}$, to the spin carried
	by the protons and the neutrons.  
	The uncertainties arising from 
        electroweak and QCD scale physics are incorporated 
        in the factors $a_p$ and $a_n$. 
	The nuclear matrix element ${\cal M}$ in 
Eq.~(\ref{Definitions.matrix.element}) is often related to the
	matrix element of the nuclear magnetic moment, which
        also consists of 
	the  matrix elements of the total proton and
        neutron spin operators:
\begin{equation}
\label{Definitions.magnetic.moment}
\mu =  \langle A\vert g_n^s {\bf S}_n + g_n^l {\bf L}_n +
	g_p^s {\bf S}_p + g_p^l {\bf L}_p\vert A\rangle.
\label{eq:13}
\end{equation}
	The {\it free particle} $g$-factors (gyromagnetic ratios) 
	are (in nuclear magnetons):
$g_n^s = -3.826$, $g_n^l = 0$, $g_p^s = 5.586$, $g_p^l = 1$. 
        The nuclear magnetic moment $\mu$ is often used as a benchmark 
        for the accuracy of 
	the calculation of ${\bf S}_p \,$ and ${\bf S}_n \,$ 
\cite{Ressell:1993qm,Ressell:1997kx}.  
        For the most interesting
	isotopes either $\langle{\bf S}^A_{p}\rangle$ 
	or $\langle{\bf S}^A_{n}\rangle$ dominates
	($\langle{\bf S}^A_{n(p)}\rangle \ll \langle{\bf S}^A_{p(n)}\rangle$).
	See, for example, 
Table~\ref{Nuclear.spin.main.table}.
\begin{figure}[ht!] 
\begin{picture}(100,150)
\put(7,0){\includegraphics{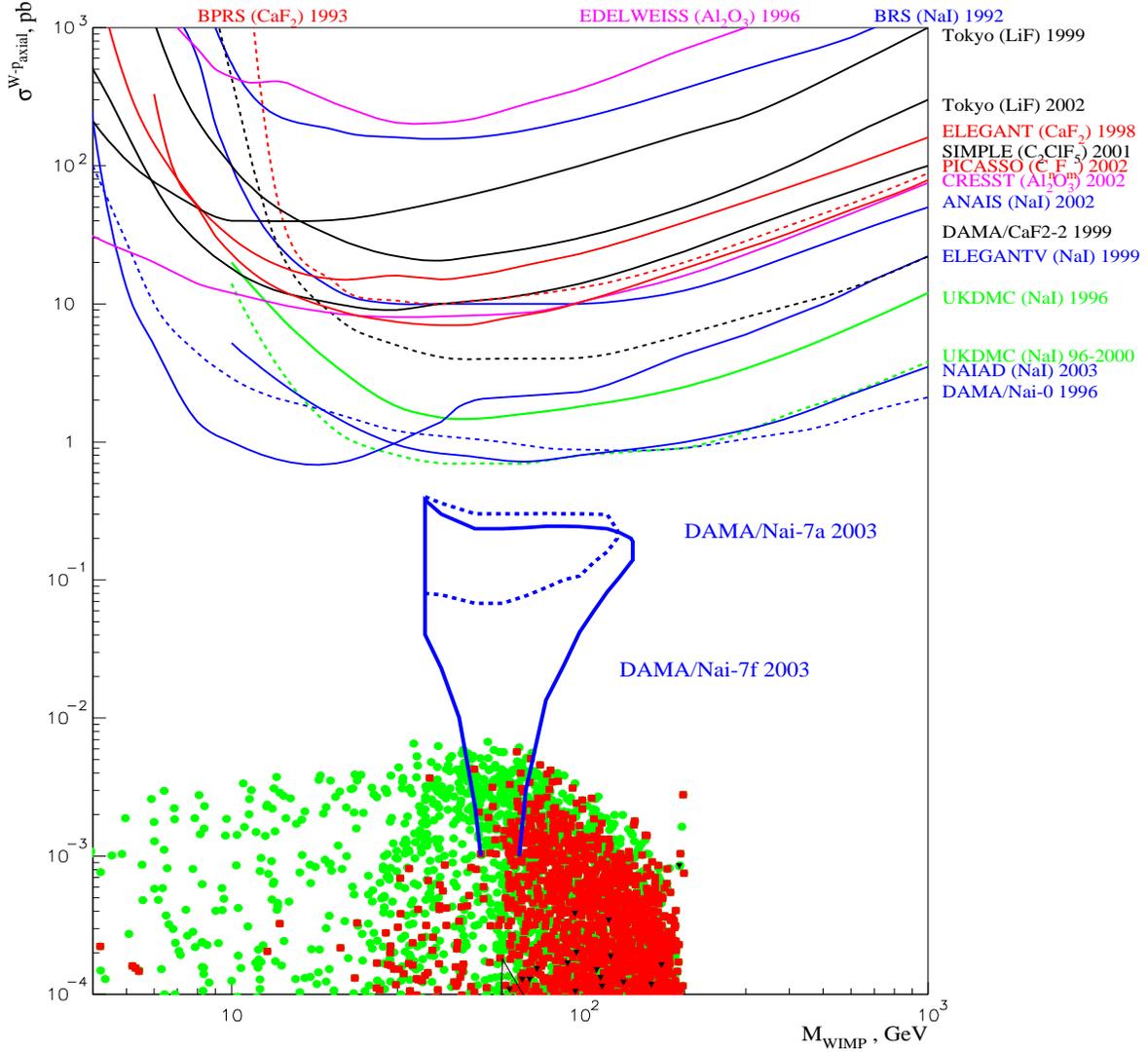}}
\end{picture}
\caption{Exclusion curves for the
        spin-dependent WIMP-{\bf proton} cross section
        ($\sigma^{p}_{{\rm SD}}$ as a function of the WIMP mass).
	DAMA/NaI-7a(f) contours 
	for the WIMP-proton SD interaction in $^{127}$I are obtained  
	on the basis of the positive 
	signature of {\em annual signal modulation}\ 
\cite{Bernabei:2003za,Bernabei:2001ve}. 
        The scattered points are calculations of
\cite{Bednyakov:2004be}.}
\label{Spin-p} 
\end{figure} 

\begin{figure}[ht] 
\begin{picture}(100,100)
\put(13,-8){\includegraphics{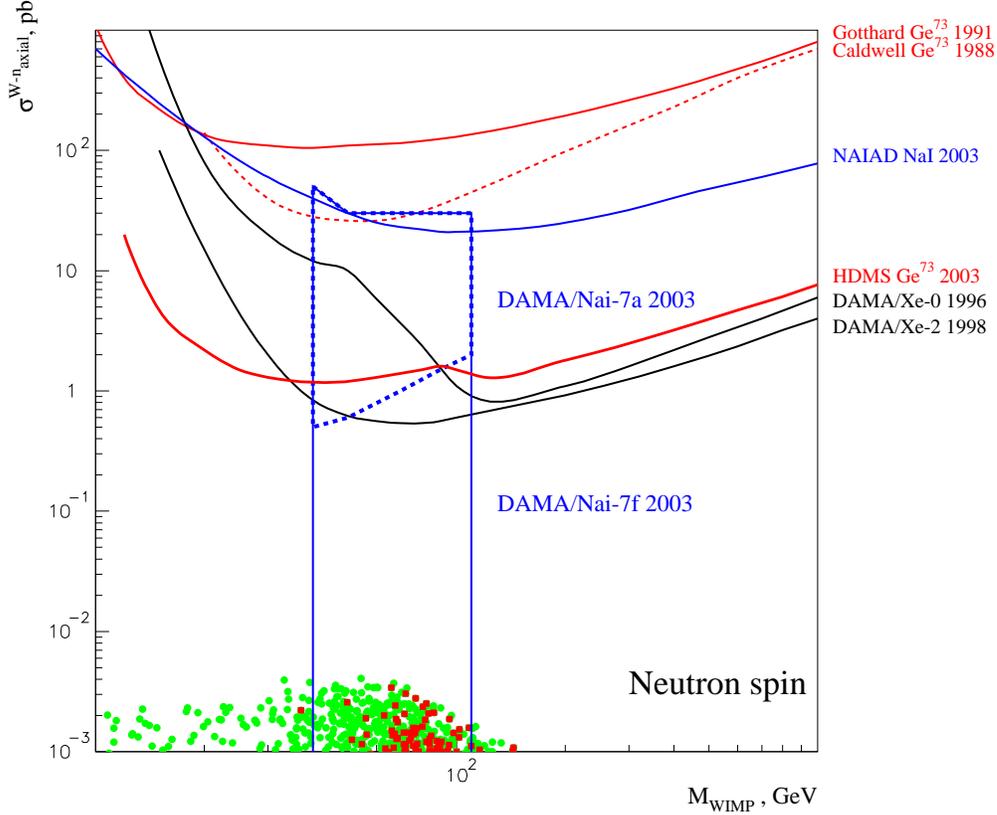}}
\end{picture}
\caption{
        Exclusion curves for the SD WIMP-{\bf neutron} cross section
        ($\sigma^{n}_{{\rm SD}}$ versus WIMP mass).}
\label{Spin-n} 
\end{figure} 

        From 
Eqs.~(\ref{NuclCS1}) 
        one can conclude             
        the spin observables in DM search 
	give us TWO independent constraints on a SUSY model via
$\sigma^{p}_{{\rm SD}}(0)$ and $\sigma^{n}_{{\rm SD}}(0)$, 
	or, equivalently, via ${a}^{}_{p}$  and ${a}^{}_{n}$.
        These constraints are usually presented in the form
	of exclusion curves obtained with different target nuclei 
(Figs.~\ref{Spin-p} and \ref{Spin-n}).
	There is only one 
        similar constraint from spin-independent DM search experiments 
(Eq.~(\ref{NuclCS0})).
	This presentation is a bit obsolete
\cite{Bernabei:2003za,Bernabei:2001ve,Bednyakov:2004be}, 
         but it allows one to compare sensitivities of different
         experiments.

\section{\boldmath Long-tail $q$-behaviour due to the spin}
\begin{figure}[!h] 
\begin{picture}(100,70)
\put(30,80){\includegraphics{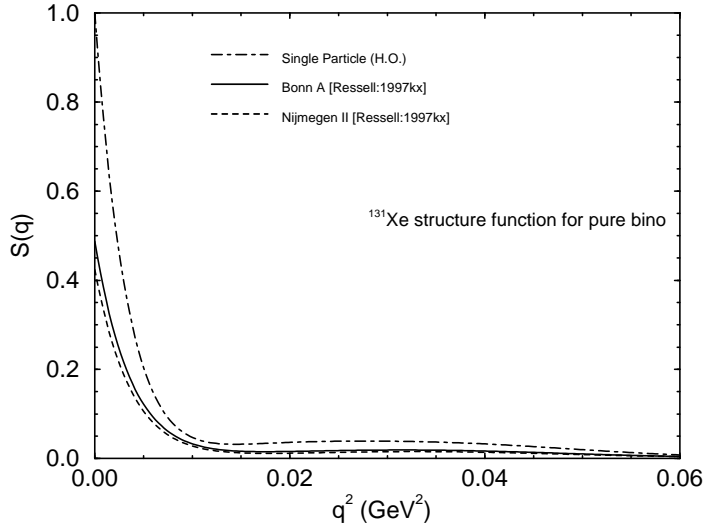}}
\end{picture} 
\caption{The $^{131}$Xe structure function for a pure bino neutralino. 
	The single-particle structure function is 
	normalized to $S(0) = 1$.  From 
\protect\cite{Ressell:1997kx}.}
\label{Xe-SF}
\end{figure}

	As $m_{\chi}$ becomes larger,
	the finite momentum transfer limit
	must be considered for heavier mass $M_A$ nuclei.  
	The differential SD event rate 
	with structure function $S^A_{\rm SD}(q)$ 
	(\ref{Definitions.spin.decomposition})
	has now the form
\begin{eqnarray}
\label{Definitions.spin.differential.rate}
\frac{dR^A_{\rm SD}}{dq^2}\!
&=&\!{\rho\over{m_{\tilde \chi}m_A}} \int v dv f(v) 
		  {{8 G_F^2}\over{(2J + 1) v^2}} S^A_{\rm SD}(q).
\end{eqnarray}
	Comparing this formula with the observed recoil spectra 
	for different targets (Ge, Xe, F, NaI, etc)
	one can directly and simultaneously 
	restrict both isoscalar and isovector 
	neutralino-nucleon effective couplings $a_{0,1}$.
	These constraints will impose most 
	model-independent restrictions on the MSSM parameter space.
	Another attractive feature of the SD 
	WIMP-nucleus interaction is the $q$-dependence of 
	SD structure function 
(\ref{Definitions.spin.decomposition}). 
	The ratio of SD to SI rate in the $^{73}$Ge detector 
	grows with the WIMP mass
\cite{Bednyakov:2000he,Bednyakov:2002mb}. 
	The growth is much greater for heavy target isotopes like xenon.
	The reason is the different behavior of 
	the spin and scalar structure functions with 
	increasing momentum transfer. 
	For example, the xenon SI structure function
	vanishes for $q^2\approx 0.02$~GeV, but 
	the SD structure function is a non-zero constant in the region
(Fig.~\ref{Xe-SF}).    
	As noted by Engel in 
\cite{Engel:1991wq},  
	the relatively long tail of the SD structure function is 
	caused by nucleons near the Fermi surface, which do the 
	bulk of the scattering. 
	The core nucleons, which dominate the SI nuclear coupling, 
	contribute much less at large $q$.
	Therefore the SD efficiency for detection of a
	DM signal is higher than the SI efficiency, 
	especially for very heavy neutralinos.

\section{One does not miss a DM signal due to the spin}
	To estimate the DM detection rate we traditionally 
	use the so-called effective scheme of MSSM (effMSSM) 
	whose parameters are defined directly at the electroweak scale, 
	relaxing completely constraints following from 
	any unification assumption
(see, for example   
\cite{Bednyakov:1998is}--\cite{Ellis:2003eg}).
	Our MSSM parameter space is determined by the entries of the mass 
	matrices of neutralinos, charginos, Higgs bosons, 
	sleptons and squarks. 
	The relevant definitions can be found in 
\cite{Bednyakov:1999vh}.
	We have included the current experimental 
	upper limits on sparticle and Higgs masses
	from the Particle Data Group. 
	Also, the limits on the rare $b\rightarrow s \gamma$ decay 
	have been imposed. 
	For each point in the MSSM parameter space (MSSM model) 
	we have evaluated the relic density of the light neutralinos
	$\Omega_{\chi} h^2_0$ with our code 
\cite{Bednyakov:2002dz,Bednyakov:2002js,Bednyakov:2002ng} based on 
\cite{Gondolo:2000ee}, 
	taking into account all coannihilation channels with 
	two-body final states that can occur between neutralinos, charginos,
	sleptons, stops and sbottoms.
	We assume $0.1< \Omega_\chi h^2  < 0.3$ 
	for the cosmologically interesting region and
	we also consider the WMAP reduction of the region
	to $0.094< \Omega_\chi h^2  < 0.129$ 
\cite{Spergel:2003cb,Bennett:2003bz}.

\begin{figure}[h] 
\begin{picture}(100,90)
\put(-10,-83){\includegraphics{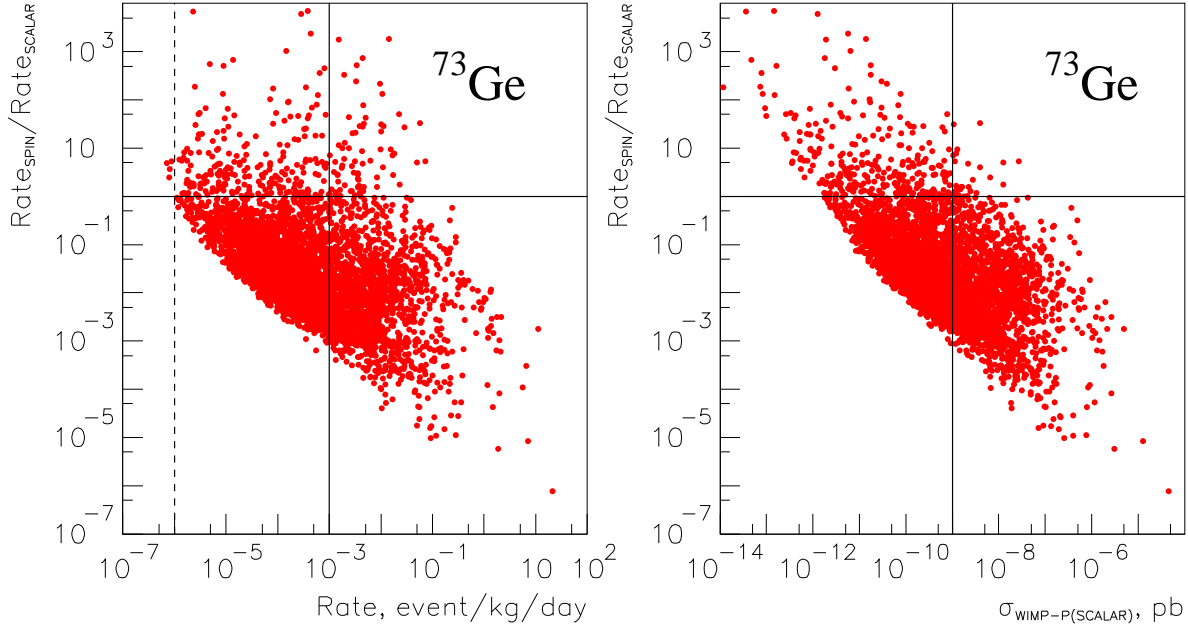}}
\end{picture}
\caption{\small 
	Ratio of the SD (spin) event rate to the   	
	SI (scalar) event rate in the $^{73}$Ge isotope
	(spin$\,=9/2$) as a function of the total (SD$+$SI) 
	event rate (left) and the scalar 
	cross section of the neutralino-proton interaction (right). 
	The solid vertical lines give the expected sensitivity of 
	one of the best future projects GENIUS
\protect\cite{GENIUS}. 
	In the region above the horizontal line the spin contribution 
	dominates.
\label{SpinandScalar}}	
\end{figure}

        From 
Fig.~\ref{SpinandScalar}
        one sees that the SD contribution
	obviously dominates in the domain of large expected rates 
	in the non-zero-spin germanium detector
	($R>0.1\,$event$/$day$/$kg). 
	But as soon as the total rate drops down to
	$R<0.01\,$event$/$day$/$kg or, equivalently, the
	SI (scalar) neutralino-proton cross section becomes
	smaller than $10^{-9}$--$10^{-10}\,$pb,
	the SD interaction may produce
	a rather non-negligible contribution to the total event rate.
	Moreover, if the scalar cross section further decreases
	($\sigma < 10^{-12}\,$pb), it becomes obvious that 
	the spin contribution alone saturates 
	the total rate and protects it from decreasing below
	$R\approx 10^{-6}$--$10^{-7}\,$event$/$day$/$kg
\cite{Bednyakov:2000vh}.
	With only a spinless detector  
	one can miss a signal caused by SD interaction.  
	An experiment 	
	aimed at {\em detecting}\ dark matter
	with sensitivity higher than $10^{-5}\,$event$/$day$/$kg
	should have a non-zero-spin target. 
	Indeed, while the scalar cross sections governed 
	mostly by Higgs exchange can be rather small, 
	the spin cross section cannot be arbitrarily small 
	because the mass of the $Z$ boson
\cite{Bednyakov:2004be}, which makes the dominant 
	contribution, is well defined, provided one ignores any 
	possible fine-tuning cancellations.
	Therefore, if an experiment with sensitivity 
	$10^{-5}$--$10^{-6}\,$event$/$day$/$kg
	fails to detect a dark matter signal,
	an experiment with higher sensitivity 
	should have a 
	non-zero-spin target and  
	will be able to detect dark matter particles 
	only due to the spin neutralino-quark interaction.

\section{Conclusion}
        There are at least three reasons to think that 
	spin-dependent 
	interaction of the DM WIMPs with nuclei
        could be very important. 
        First, contrary to the only one constraint for SUSY models available 
        from the spin-independent 
	WIMP-nucleus interaction, the SD WIMP-nucleus 
        interaction supplies us with two such constraints.
	Second, for heavy target nuclei and heavy WIMP masses
	the SD efficiency to detect a DM signal 
        is much higher than the SI efficiency.  
        Finally, the absolute lower bound for the DM detection rate 
	can naturally be due to SD interaction.
	An experiment aimed at {\em detecting}\ DM 
	with sensitivity higher than $10^{-5}\,$event$/$day$/$kg
	should have a non-zero-spin target.

\smallskip
      The authors 
      have a pleasure to thank Prof. H.V.Klapdor-Kleingrothaus
      for fruitful and permanent cooperation as well as 
      the organizers 
      of the Baldin Seminar for their invitation to give  
      this pedagogical review talk whose subject is 
      beyond the traditional scope of the Seminar.

\end{document}